\documentclass[11pt]{article}
\usepackage{moriond,epsfig}

\def\Journal#1#2#3#4{{#1} {\bf #2}, #3 (#4)}

\newcommand{\bi}[1]{\mbox{\boldmath ${#1}$}}

\newcommand{\gtrsim}{\mbox{\raisebox{2.7pt}{$>$}\hskip -9pt
\raisebox{-2.6pt}{$\sim$}}}

\newcommand{\eref}[1]{(\ref{#1})}

\newcommand{\e}{{\rm e}}
\newcommand{\rmd}{{\rm d}}
\newcommand{\rmi}{{\rm i}}

\newcommand{\half}{{\textstyle{\frac{1}{2}}}}

\newcommand{\up}{\uparrow}
\newcommand{\dn}{\downarrow}

\newcommand{\al}{\alpha}

\newcommand{\de}{\delta}

\newcommand{\eps}{\epsilon}

\newcommand{\om}{\omega}
\newcommand{\Om}{\Omega}

\newcommand{\brkt}[1]{\left({#1}\right)}
\newcommand{\sqbrkt}[1]{\left[{#1}\right]}
\newcommand{\anbrkt}[1]{\left<{#1}\right>}
\newcommand{\stbrkt}[1]{\left|{#1}\right|}

\newcommand{\bra}[1]{\left<{#1}\right|}
\newcommand{\ket}[1]{\left|{#1}\right>}

\begin{document}
%%%%%%%% HAS BEEN SUPPRESSED \vspace*{4cm}
\title{DOES BERRY PHASE EXIST FOR A SYSTEM COUPLED TO ITS ENVIRONMENT?}

\author{\underline{Robert S. Whitney}$^1$, Yuval Gefen$^1$}

\address{$^1$ Department of Condensed Matter Physics,
Weizmann Institute of Science, Rehovot 76100, Israel.}

\maketitle\abstracts{
Berry phase was originally defined for systems whose states
are separated by finite energy gaps.
One might naively expect that a system 
without a gap cannot have a Berry phase.
Despite this we ask whether a Berry phase
can be observed in a system which has a continuous spectrum
because its coupling to the environment has broadened its energy levels.  
We find that, contrary to the above naive expectation, there are conditions 
under which the Berry phase is observable.  However it is modified
by the presence of the environment and no longer has a simple geometric
interpretation.
The model system we consider is a spin-half in a slowly rotating
magnetic field, with the spin also coupled to a Ohmic 
environment of harmonic  oscillators (spin-boson model).  Here we discuss the 
high-temperature limit of this model.
We then interpret our results in terms of a spin under the influence of a 
classical stochastic field.
}

\section{Motivation and Summary}\label{sect10}

Originally Berry phase \cite{Berry84} was defined for systems whose states
were separated by finite energy gaps.
Here we ask whether a Berry phase
can be observed in a system whose spectrum is continuous,
more specifically: a system which is not completely
isolated from its environment.
All real systems are coupled, at least weakly, to their environment and as a
result never have a truly discrete energy level spectrum.
The naive argument would be that, since the spectrum of the system does not
have a gap, the parameters in the Hamiltonian could never be varied slowly
enough to be considered adiabatic.  This in turn would mean one could not
observe a Berry phase in such a system.
If this were the case one could never observe a Berry phase.

The above argument must be too naive: experiments have been
carried out which do observe the Berry phase, both directly and
indirectly \cite{Anandan97,Thoulessbook,Ao93Zhu97}. 
We therefore take a simple model
in which a quantum system which exhibits a Berry phase is coupled
to many other quantum degrees of freedom. We then ask two
questions. Firstly, under what conditions can the Berry phase be
observed? Secondly, is the observed Berry phase the same as that
of the isolated system, and if not is it still geometric in
nature? While others have investigated systems with Berry phases
coupled to other degrees of freedom
\cite{Gaitan98,Ao99,Avron98Avron99}, we believe we are the first
to explicitly ask these two questions.

We should make it clear that we distinguish between the system and the
environment in the following way.  
{\it The system} is something which we have
experimental control over.
Thus we can prepare the system in any state we like
and choose all parameters in the Hamiltonian under which it evolves.
{\it The environment} consists of all the degrees of freedom over which
we have little experimental control.
Usually the most we can do to the environment is to ensure
the universe (system $+$ environment)
is in thermal equilibrium, with a temperature $T$.
If we are able to
take $T$ to zero, then
we can prepare the universe in its ground state. 
However to measure a Berry phase, we must devise a procedure for
measuring the phase evolution of this state and getting rid of the
dynamic part of the phase.
Most such procedures involve the mixing of a large number of
eigenstates (of the universe), and hence will lead to dephasing as is
discussed below \cite{Whitney01}.
This implies that the extremely interesting recent work of
Avron and \hbox{Elgart \cite{Avron98Avron99}} --- on adiabatic evolution
of eigenstates of models similar to ours --- is not directly relevant to
our work.

We choose to investigate a spin-half which is coupled to both
a magnetic field and an environment.
When isolated from its environment,
this spin exhibits a Berry phase if we slowly rotate the magnetic field
around a closed loop.
We model the environment as a bath of harmonic oscillators coupled
to the spin.
This model has been chosen for its simplicity, despite this it is relevant to
recent suggestions for using Berry phases to control the qubits in a
quantum computer \cite{Jones00Ekert00,Falci00}.
This is of interest because the coupling to the environment is significant
for most physical realisations of qubits.
While we make no attempt to accurately model the real coupling between
the qubit and its environment, we believe our results give an indication of
what to expect in the real system.

In this article we concentrate on the case where the frequencies of the
oscillators have been chosen so that the {\it bath is Ohmic}
\cite{Caldiera83},
and these oscillators are initially at a {\it high-temperature}.
We find that coupling the spin-half to this type of environment causes
four effects.
(i) Exponential decay to a state in which the spin is thermalised with
the bath, the time-scale associated with this is the spin flip time, $T_1$.
(ii) Exponential decay of observables containing phase information,
the time-scale associated with this is the dephasing time, $T_2$.
(iii) A Lamb shift \cite{Schweber61} of the energy-levels of the spin by an
amount, $\de E$.
(iv) A modification to the Berry phase, $\de \Phi_{\rm Berry}$.
All four of these effects go like the second power of the coupling between the
spin and its environment, the functional form of the results are given in
Eq. \eref{50}.  Of these (ii) and (iv) are of most interest to us.
Effect (ii) tells us when we can measure the Berry (or any other)
phase,  while effect (iv) tells us what the Berry phase we measure will
be.

Effect (ii) means that one cannot perform an arbitrarily long
experiment to measure a phase.  Even the most sensitive experiment
will not be able to measure the phase if the time of the
experiment is much larger than $T_2$, because of the exponential
decay of the observables containing the phase information.  This
constraint on the time of the experiment is particularly relevant
for Berry phase experiments, since they must be carried out over
long times to ensure adiabaticity. So what is the Berry phase for
an experiment in which the parameters of the Hamiltonian are taken
around a closed loop in a finite time, $t_{\rm period}$? The Berry
phase is the component of the phase which is independent of $t_{\rm period}$.
Other contributions to the total phase are the dynamic phase which
scales linearly with with $t_{\rm period}$, and non-adiabatic
contributions which are proportional to $t_{\rm period}$ to some
negative power.  Clever experiments (see Section \ref{sect50}) can
remove the dynamic phase but not the non-adiabatic contributions.
One can argue that the Berry phase is present even for very fast
experiments, it is just that then it is masked by the
non-adiabatic contributions to the phase. For the Berry phase to
be accurately measured it must be much larger than the
non-adiabatic contributions to the total phase. This can only be
achieved by carrying out the experiment slowly enough.

Let us assume that during an experiment on a spin-half in a
magnetic field, $\bi{B}(t)$, we slowly change the
direction of the field while keeping its magnitude fixed.  Then at
all times the energy gap between the up and down state is $g \mu
B$, where $B= \stbrkt{\bi{B}(t)}$ is time-independent, $g$ is the
Land\'e g-factor, and $\mu \equiv {e \hbar \over 2mc}$ is the Bohr
magneton. In this case the non-adiabatic contribution to the phase
accumulated over a time, $t$, is of order $\hbar/(g\mu Bt)$. To
measure the Berry phase --- which is of order one --- the
non-adiabatic contributions to the total phase must be much less
than one. Thus the time of the experiment should satisfy $t_{\rm
period} \gg \hbar/(g\mu B)$. If dephasing occurs on a time-scale
$T_2$ then we cannot measure the total phase for times larger than
$T_2$. Unless $g\mu B T_2/\hbar \gg 1$ the non-adiabatic
contributions to the total phase in an experiment which takes time
$T_2$ will not be small enough for us to accurately measure the
Berry phase.

Now that we understand the conditions under which we can measure
the Berry phase, we ask if the Berry phase is purely geometric.
The answer is no. Effect (iv) tells us that in general there are
environment-induced contributions to the Berry phase which are not
geometric in nature. Remember we are choosing to define the Berry
phase as any contribution to the total phase which is independent
of the time, $t_{\rm period}$, that it takes to vary the
parameters of the system Hamiltonian around a closed loop. Is this
non-geometric contribution to the Berry phase observable?  It can
be very large, but only if the coupling to the environment is
large, in which case the experiment dephases quickly. Then a Berry
phase measurement carried out before the system dephases will have
significant non-adiabatic contributions to the total phase
 which may mask $\de \Phi_{\rm Berry}$.
Therefore the environment-induced contribution to the Berry phase can
only be observed if
\begin{eqnarray}
\de \Phi_{\rm Berry} \gtrsim {\hbar \over g\mu BT_2} .
\end{eqnarray}
Only then is $\de \Phi_{\rm Berry}$ of the same order or larger
than the non-adiabatic corrections to the phase that occur when
the experiment is carried out in a time $t_{\rm period}\sim T_2$.
If we look at the functional form of $T_2$ and $\de \Phi_{\rm
Berry}$ in \eref{50} we see that there are a wide range of
parameters for which $\de \Phi_{\rm Berry} \sim {\hbar \over g\mu
BT_2}$.  In this parameter range the environment-inducted correction to
the Berry phase is just about observable in an experiment which
takes a time $t_{\rm period}\sim T_2$.

The remainder of this article is arranged as follows.
In Section \ref{sect40} we discuss the model and summarise our results
for a high-temperature Ohmic bath.
Then in Section \ref{sect60} we describe the
high-temperature Ohmic environment as a randomly-fluctuating magnetic
field, and thus develop a more intuitive understanding of one limit of this
model.
To do this we first give a brief summary
of Berry phase for an isolated spin (Section \ref{sect50}).

\section{The model and high-temperature results} \label{sect40}

We will discuss the details of the derivation of the results elsewhere
\cite{Whitney01}; here we give a very brief overview of the derivation.
We start with the following Hamiltonian,
\begin{eqnarray}
{\cal H} = -{g \mu \over 2} \bi{B}(t) \cdot \hat{\bi{\sigma}}
-{g\mu \over 2}\sum_i C \hat{q}_i \hat{\sigma}_z
+ \sum_i \sqbrkt{{\hat{p}_i^2 \over 2m} + {m \Om_i^2 \hat{q}_i^2
\over 2}} .
\end{eqnarray}
It is a spin-boson model \cite{Leggett87} with a time-dependent field, 
\bi{B}(t).
We consider an Ohmic bath of oscillators with an upper frequency cut-off,
$\Om_{\rm max}$, thus
\begin{eqnarray}
\sum_i \brkt{\cdots}_{\Om_i}
= \al \int_0^\infty \rmd \Om \ \Om^2 \e^{-\Om/\Om_{\rm max}} \brkt{\cdots}_\Om
\end{eqnarray}
We write the spin operators as operators on a two fermionic system
$\sigma_z = f^\dag_1 f_1 - f^\dag_2 f_2$,
$\sigma_+ = f^\dag_1 f_2$ and
$\sigma_- = f^\dag_2 f_1$,
we then write the Feynman propagator for the system as a coherent-states path
integral in which the fermionic degrees of freedom represented by Grassman
variables.
We integrate out the oscillators exactly \cite{Feynman63,Caldiera83}.
The effective action of the resulting path integral contains a term which is
quartic in the
fermionic degrees of freedom and couples the state of the spin at time $t_1$
to the state of the spin at time $t_2$.
We carry out a perturbation expansion of this quartic term.
We then use the fact that
for an Ohmic bath the coupling between the spin state at $t_1$ and $t_2$
is dominated by $t_1-t_2 \sim \Om_{\rm max}^{-1}$.
Thus in the limit where the time of the experiment $t \gg \Om_{\rm max}^{-1}$,
the perturbation expansion is dominated by a certain class of diagrams
that we can sum to all orders.

Here we only discuss results for a {\it high-temperature} Ohmic environment.
By high-temperature we mean that $k_{\rm B}T \gg \hbar \Om_{\rm max}$,
other limits are discussed in reference 8.
For concreteness we take the case where the field 
$\bi{B}(t)= B \brkt{
\hat{\bi{x}}\sin \theta \cos \om t + 
\hat{\bi{y}}\sin \theta \sin \om t + \hat{\bi{z}} \cos \theta}$
in an experiment which takes a time $t_{\rm period}=2\pi/\om$.
The results --- written in terms of
the dimensionless coupling between the spin and the oscillators
$\tilde{C} = g\mu C\brkt{\al/m\hbar}^{1/2}$ --- are as follows,
\begin{eqnarray}
{1\over T_1} &=& {\pi \tilde{C}^2(k_{\rm B}T) \over 2\hbar}
\e^{-\gamma} \sin^2 \theta
\qquad \qquad \qquad \qquad
{1\over T_2} \ =\
{\pi \tilde{C}^2(k_{\rm B}T) \over 4\hbar} \brkt{2 \cos^2 \theta
+\e^{-\gamma} \sin^2 \theta}
\nonumber \\
{\de E  \over \hbar} &=& {\tilde{C}^2 (k_{\rm B}T)\over 4\hbar}
f(\gamma) \sin^2\theta \qquad \qquad \quad \quad \ \de \Phi_{\rm Berry} \ =\
{\pi \tilde{C}^2(k_{\rm B}T) \over 2\hbar \Om_{\rm max}}
 f'(\gamma)\sin^2 \theta \cos \theta \label{50}
\end{eqnarray}
where $\gamma=g\mu B/\hbar \Om_{\rm max}$. 
The function $f(x)=\e^x {\rm Ei}(-x) - \e^{-x}{\rm Ei}(x)$
where we define ${\rm Ei}(x)$ as the principal-value of the
Exponential integral, $\int_{-x}^\infty \rmd t \e^{-t}/t$.
Hence
$f'(x)\equiv{\rmd \over \rmd x}f(x)=\e^x {\rm Ei}(x)+\e^{-x}{\rm Ei}(-x)$.

\section{Berry phase for an isolated spin : Rotating reference frame trick}
\label{sect50}

Consider a spin-half coupled to a magnetic field, $\bi{B}(t)$, via
the Hamiltonian,
\begin{eqnarray}
{\cal H}=-{g \mu \over 2}\bi{B}(t) \cdot \hat{\bi{\sigma}}
\end{eqnarray}
Now imagine that the magnetic field rotates with angular velocity,
$\bi{\om}(t)$, so that
${\rmd \over \rmd t} \bi{B}(t) = \bi{\om}(t)\times\bi{B}(t)$.
We can go to a reference frame which rotates with the $\bi{B}$-field.
In this new non-inertial frame the spin experiences
a pseudo-field, as well as the time-independent field $\bi{B}'= \bi{B}(0)$.
The pseudo-field is simply $-\hbar \bi{\om}(t)/g\mu$.
Thus all the time-dependence in the problem has been moved from $\bi{B}$
to $\bi{\om}$.
In the rotating frame the Hamiltonian is
\begin{eqnarray}
{\cal H}'=-{g \mu \over 2}
\sqbrkt{\bi{B}' - {\hbar \over g \mu}\bi{\om}(t)}
\cdot \hat{\bi{\sigma}} .
\end{eqnarray}
Note that if we transform the initial state to the rotating frame, evolve it
using ${\cal H}'$, then transform back to the non-rotating frame, we end up
with a final state $\big|\psi_f'\big>$.
Careful analysis shows that the true final state is actually
$\big|\psi_f\big>= \e^{\rmi \phi \hat{\sigma}_z}\big|\psi_f'\big>$,
however in all
cases we will consider $\phi$ is a multiple of $\pi$.  The resulting
sign difference between  $\big|\psi_f\big>$ and $\big|\psi_f'\big>$ will
be of no consequence and can be ignored.

The power of transforming to the rotating frame becomes clear if we consider
a rotation of the $\bi{B}$-field for which $\bi{\om}$ is time-independent.
Then the transformation turns a time-dependent problem
into a time-independent one.  The time-independent problem can then be solved
by working in the basis where ${\cal H}'$ is diagonal.

We consider the following experiment to measure the Berry phase.
We start with a $\bi{B}$-field, $\bi{B}(0)= B \brkt{\hat{\bi{x}}\sin
\theta+\hat{\bi{z}} \cos \theta}$, which we then rotate with
angular velocity $\bi{\om}= \hat{\bi{z}}\om$ for a time $t_{\rm
period}=2\pi/\om$. In the limit $\om \to 0$, the component of the
spin-state which is initially up relative to $\bi{B}(0)$ acquires a phase 
$\Phi = \half \Theta +\int_0^{t_{\rm period}} E_{\up}(t) \rmd t $,
while the state which is initially down acquires a phase equal to $-\Phi$. 
The first term in
$\Phi$ is the Berry phase (for a spin-half), $\Phi_{\rm Berry} =
\half \Theta$, where $\Theta = 2 \pi \brkt{1-\cos \theta}$ is the
solid-angle enclosed by the path the $\bi{B}$-field takes in the
non-rotating frame. The second term is the dynamic phase,
$\Phi_{\rm dyn}= \int_0^{t_{\rm period}} E_{\up}(t) \rmd t$, where
$E_{\up}(t)= \half g\mu B$ is the energy of the instantaneous
eigenstate at time $t$.

We are in the adiabatic limit, which means that the dynamic phase
is much larger than the Berry phase.  If we wish to carry out an
experiment which yields the Berry phase, rather than the sum of the Berry
and dynamic phases, then we can eliminate the dynamic phase using
the following spin-echo technique \cite{Jones00Ekert00}.
We (a) adiabatically rotate the $\bi{B}$-field around a closed loop, 
then (b) flip the spin, 
(c) rotate the field in the opposite direction round the same closed loop, 
and (d) flip the spin again.
By flip the spin, we mean $\ket{\up} \leftrightarrow \ket{\dn}$
where the up and down states
are defined relative to the direction of $\bi{B}$-field at that time.
This can be achieved by applying a fast $\pi$-pulse in a direction
perpendicular to the $\bi{B}$-field.
When we do (a)-(d) the Berry phases of the first and second loop add up,
while the dynamic phases cancel.  Thus the phase acquired by the spin is
purely dependent on the Berry phase round the loop.

Thus we will consider a field, $\bi{B}(t)$, which rotates with
angular velocity $\bi{\om}= \hat{\bi{z}}\om$ for $0<t<t_{\rm
period}$ (where $t_{\rm period}= 2 \pi /\om$) and $\bi{\om}=
-\hat{\bi{z}}\om$ for $t_{\rm period}<t<2t_{\rm period}$, with a
spin flip at time $t_{\rm period}$ and time $2t_{\rm period}$.
Then the probability of an initial state $\ket{u_0,v_0} = u_0
\ket{\up}+v_0\ket{\dn}$ ending up in a state $\ket{u,v}$ at time
$2t_{\rm period}$ is
\begin{eqnarray}
P\brkt{\ket{u,v},2t_{\rm period};\ket{u_0,v_0}} &=&
\stbrkt{\bra{u,v} \hat{\sigma_x} \exp \sqbrkt{\rmi
\Phi(-\om)\hat{\sigma}_z} \hat{\sigma_x} \exp \sqbrkt{\rmi
\Phi(\om)\hat{\sigma}_z}\ket{u_0,v_0}}^2
\nonumber \\
&=&
u^*uu_0^*u_0
+ v^*u v_0^*u_0 \e^{\rmi 4\Phi_{\rm Berry}}
+ u^*v u_0^*v_0 \e^{-\rmi 4\Phi_{\rm Berry}}
+ v^*v v_0^*v_0
\label{115}
\end{eqnarray}
where all spin-states are defined relative to the axis
of the field $\bi{B}(0)$.

\section{Berry phase and decoherence in a randomly fluctuating magnetic field}
\label{sect60}

%%%%%%%%%%%%%%%%%%%%%%%%%%%%%%%%%%%%%%%%%%%%%%%%%%%%
\begin{figure}
\centerline{\psfig{figure=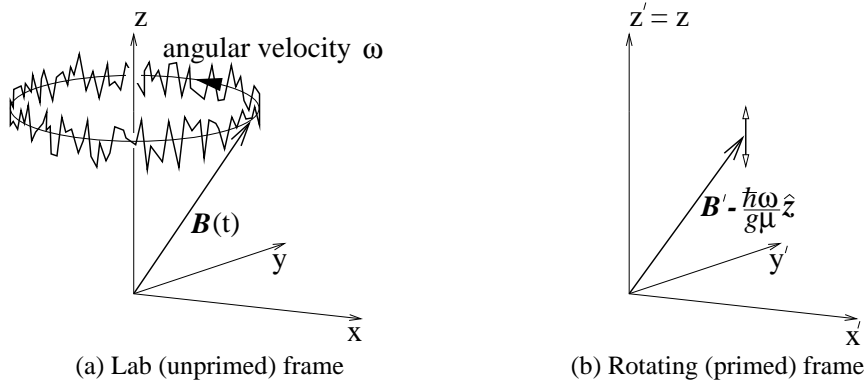,height=2.0in}}
\caption{Here we show the magnetic field discussed in Section \ref{sect60},
in both the laboratory frame (a) and the rotating frame (b).
The smooth line in (a) is the time-dependence of the field, $\bi{B}(t)$,
while the irregular line represent the total field, 
$\brkt{\bi{B}(t)+\hat{\bi{z}}K(t)}$.  It fluctuates wildly because $K(t)$
fluctuates randomly with time.
In the rotating frame (b) both the field $\bi{B}'$ and the  pseudo-field 
$\brkt{-\hat{\bi{z}} \hbar \om /g\mu}$ are time-independent, 
while the field $K(t)$ fluctates. 
The vertical double-ended arrow represent the fluctuations
of the total field, 
$\brkt{\bi{B}'- \hat{\bi{z}}\hbar \om / g \mu + \hat{\bi{z}}K(t)}$.
\label{fig10}}
\end{figure}
%%%%%%%%%%%%%%%%%%%%%%%%%%%%%%%%%%%%%%%%%%%%%%%%%%%%

A spin in an experimentally controlled magnetic field $\bi{B}(t)$
with a $z$-axis coupling to a high-temperature Ohmic bath of oscillators
is equivalent to  the spin in a field  $\bi{B}(t)+\bi{K}(t)$
where $\bi{K}(t)=\hat{\bi{z}}K(t)$ is a randomly-fluctuating field 
\cite{Caldiera83}.
The Hamiltonian for this model is simply
${\cal H}=-{g\mu \over 2}\sqbrkt{\bi{B}(t)+\hat{\bi{z}}K(t)} 
\cdot \hat{\bi{\sigma}}$.
If there is an upper cut-off on the frequencies
of the oscillator, $\Om_{\rm max}$, then $K(t)$ is
Gaussianly distributed and uncorrelated for
times longer than  $\Om_{\rm max}^{-1}$.
Thus on time-scales much greater than $\Om_{\rm max}^{-1}$ we can average over
$K(t)$ using,
\begin{eqnarray}
\anbrkt{\cdots}
={\int  \sqbrkt{{\textstyle \prod_{j=0}^{N-1}}\rmd K_j}
(\cdots) \exp \sqbrkt{-{\eps \over 2\kappa^2}\sum_{j=0}^{N-1} K_j^2} \over
\int  \sqbrkt{{\textstyle \prod_{j=0}^{N-1}}\rmd K_j}
 \exp \sqbrkt{-{\eps \over 2\kappa^2}\sum_{j=0}^{N-1} K_j^2}}
\label{120}
\end{eqnarray}
where $K_j \equiv K(j\eps)$ and the period of each
time-slice $\eps= t/N$.
We require that $\eps$ is of the order or larger than $\Om_{\rm max}^{-1}$.
This random-field model is equivalent to
the Ohmic high-temperature bath of oscillators with
\begin{eqnarray}
\kappa^2 = {\al \pi C^2 (k_{\rm B}T) \over m}
={\pi \hbar \tilde{C}^2 (k_{\rm B}T) \over g^2 \mu^2} .
\label{125}
\end{eqnarray}
In general this equivalence does not aid our calculations. However
there is one limit for which the random-field interpretation
enables us to derive the results in a much simpler and more
intuitive manner. This is the limit in which $\Om_{\rm max} \ll g
\mu B/\hbar$. Then the spin in the random field model can
adiabatically follow the field $\bi{B}(t)+\hat{\bi{z}}K(t)$, because
$K(t)$ varies smoothly on time-scales of order $\Om_{\rm
max}^{-1} \gg \hbar/(g \mu B)$. The exact requirement for the spin
to adiabatically follow the field is that at all times 
\begin{eqnarray}
{\rmd \over \rmd t}\sqbrkt{\bi{B}(t)+\hat{\bi{z}}K(t) 
\over \stbrkt{\bi{B}(t)+\hat{\bi{z}}K(t)}} 
\ll {g \mu \over \hbar} \stbrkt{\bi{B}(t)+\hat{\bi{z}}K(t)} .
\label{127}
\end{eqnarray}
We assume that
fluctuations in $K$ are given by $\big<K^2 \big>^{1/2}
\sim \kappa \Om_{\rm max}^{1/2}$ for time-scales greater than
$\Om_{\rm max}^{-1}$, while $K$ varies smoothly on time-scales
shorter than that. If $B \gg \big<K^2\big>^{1/2}$ then we do
not need any detailed information on the dynamics of $K$ on
these short time-scales.  We will restrict ourselves to the
limit where $B \gg \big<K^2\big>^{1/2}$ and $B \gg \hbar
\Om_{\rm max} /g\mu$.  Hence the adiabaticity condition, 
Eq. \eref{127}, is reduced to $\kappa \Om_{\rm max}^{3/2}/B^2 \ll 1$,
which is automatically fulfilled in this limit.
The limit defined by these approximations
is easy to work in because it is only in this limit that we are
allowed to take the adiabatic limit first and then do the
averaging.

If we consider the same time-dependence for  $\bi{B}(t)$ as we did
for the isolated spin, then we can simplify the problem
significantly by going to the reference frame which rotates with
$\bi{B}(t)$, see Fig. \ref{fig10}. 
Remember that in the non-rotating frame $K(t)$
fluctuates in the $\hat{\bi{z}}$-direction.  
Thus it unaffected by going to the rotating frame,
in other words $\bi{K}'(t)$ as $\bi{K}(t)$. In the frame which rotates
with angular velocity $\bi{\om}=\pm \hat{\bi{z}}\om$ the total phase
acquired in time, $t_{\rm period}$, is simply the dynamic phase in this 
rotating frame,
\begin{eqnarray}
\Phi(\pm \om) = {g\mu \over 2\hbar}
\int_0^{t_{\rm period}}
\stbrkt{\bi{B}'\mp(\hbar/g\mu)\hat{\bi{z}}\om+\hat{\bi{z}}K(t)}\rmd t
\label{130}
\end{eqnarray}
The path of
$\brkt{\bi{B}(0)\mp (\hbar/g\mu)\hat{\bi{z}}\om +\hat{\bi{z}}K(t)}$ encloses
no area in the rotating frame, so there is no Berry phase in this frame. 
It is important to remember that this does not mean there is no Berry phase 
in the laboratory frame.  We have already seen this for the case where
$\hat{\bi{z}}K(t)=0$.  Then there is no Berry phase in the rotating frame, 
but there is in the laboratory frame (as shown in Section \ref{sect50}). 
To identify the Berry phase we must first ensure
$\om= 2\pi/t_{\rm period}$, the Berry phase will then be the
$t_{\rm period}$-independent component of the total phase.

We now wish to calculate the averaged propagation probability for
evolution under the time-dependent $\bi{B}$-field described in Section
\ref{sect50}.  Before averaging the propagation probability is given by
\eref{115} with the phase $4\Phi_{\rm Berry}$ replaced by
$2\sqbrkt{\Phi(\om)-\Phi(-\om)}$, 
where $\Phi(\pm \om)$ are given in \eref{130}.
We can then average using \eref{120}.  While in general carrying out the
averaging is difficult, it is straightforward in the limit we are considering
--- $B \gg \big<K^2\big>^{1/2}$ and $B \gg \hbar \Om_{\rm max} /g\mu$.
We will time-slice the propagator, writing the integral over time
in the exponent as a sum of $N$ time-slices each of which takes a
time $\eps=t/N$. 
We use \eref{120} to average the propagation
probability, by evaluating a few Gaussian integrals.
We find the random fluctuations of $K(t)$ cause the phase to acquire
an imaginary part and an extra real part. The former causes an
exponential decay of all in terms containing phase information.
The latter means that the phase which is observed is different
from the $K(t)=0$ case. 
We extract the contributions to the Berry phase
by finding the parts of the total phase which 
which are independent of $t_{\rm period}$.
Similarly the contribution to the dynamic phase is the part of the total
phase which scales linearly with  $t_{\rm period}$.  We then
interpret the $\kappa$-dependent contribution to the dynamic phase as being
a Lamb shift of the energy levels.   
In the language introduced in Section \ref{sect10},
\begin{eqnarray}
{1\over T_1} &=& 0 
\qquad \qquad \qquad \qquad \qquad \qquad \qquad \quad \hskip 0.02 truein \ \
{1\over T_2} \ =\ {1 \over 2}\brkt{g\mu \kappa \over \hbar}^2 \cos^2 \theta
\nonumber \\
{\de E  \over \hbar} &=& {1 \over \#}{g \mu \kappa^2 \Om_{\rm
max}\over 2\hbar B}\sin^2 \theta 
\qquad \qquad \quad \quad\ \de \Phi_{\rm Berry}
\ = \ { 1\over \#} {\pi \kappa^2 \Om_{\rm max} \over 2B^2} \sin^2 \theta 
\cos \theta
.
\label{140}
\end{eqnarray}
where $\#= \eps\Om_{\rm max}$ is a integer larger than one but
still of order one.  This means that the results are not independent
of our choice of time-slicing for the averaging.  This is not
surprising since there is a time-scale ($\Om_{\rm max}$) below
which the field $K(t)$ is highly-correlated. This calculation is
not sophisticated enough to tell us what $\#$ should be. However
we can compare \eref{140} with \eref{50} using \eref{125} and the
large-$x$ expansion of the Exponential integral, ${\rm Ei}(x)=
-\e^x\brkt{x^{-1} + x^{-2} + \cdots}$.  In this case we see the
two sets of results agree completely if we set $\# = \pi$. 

This derivation makes it very clear that $T_1 \to \infty$ in this
limit because slow fluctuations of the field can dephase but not
flip the spin. It also shows --- 
in a more transparent manner than the calculation discussed 
in Section \ref{sect40} ---
that the environment can modify the
dynamic and Berry phases.

\section*{Acknowledgments}
We gratefully acknowledge useful discussions with Qian Niu, Rosario Fazio 
and Frank Wilhelm.
The research was supported by the U.S.-Israel Binational Science Foundation
(BSF), by the Minerva Foundation, by the Israel Science Foundation 
-- Center of Excellence, and by the German-Israel Foundation (GIF).

\end{document}